%% file: main.tex
\documentclass[fleqn,usenatbib]{rasti} 

\usepackage[T1]{fontenc}

\DeclareRobustCommand{\VAN}[3]{#2}
\let\VANthebibliography\thebibliography
\def\thebibliography{\DeclareRobustCommand{\VAN}[3]{##3}\VANthebibliography}

\usepackage{import}
\usepackage{lipsum}
\usepackage{bm}
\usepackage{dsfont}
\usepackage{color}
\usepackage{graphicx}	
\usepackage[tbtags]{amsmath}	
\usepackage{physics}
\usepackage{xcolor}

\renewcommand{\*}[1]{\bm{#1}}

\title[SBI of the sky-averaged 21-cm signal]{Simulation-Based Inference of the sky-averaged 21-cm signal from CD-EoR with REACH}

\author[Saxena et al.]{Anchal Saxena,$^{1}$\thanks{E-mail: a.saxena@rug.nl}
P.\ Daniel Meerburg,$^{1}$
Christoph Weniger$^{2}$,
Eloy de Lera Acedo$^{3, 4}$,
and \newauthor Will Handley$^{3, 4}$
\\
$^{1}$Van Swinderen Institute, University of Groningen, Nijenborgh 4, 9747 AG Groningen, The Netherlands\\
$^{2}$Gravitation Astroparticle Physics Amsterdam (GRAPPA), Institute for Theoretical Physics Amsterdam and\\
Delta Institute for Theoretical Physics, University of Amsterdam, Science Park 904, 1098 XH Amsterdam, The Netherlands\\
$^{3}$Cavendish Astrophysics, University of Cambridge, Cambridge, UK\\
$^{4}$Kavli Institute for Cosmology, Madingley Road, Cambridge CB3 0HA, UK
}

\date{Accepted XXX. Received YYY; in original form ZZZ}

\pubyear{2024}

\begin{document}
\label{firstpage}
\pagerange{\pageref{firstpage}--\pageref{lastpage}}
\maketitle

\begin{abstract}
    The redshifted 21-cm signal from the Cosmic Dawn and Epoch of Reionization carries invaluable information about the cosmology and astrophysics of the early Universe. Analyzing data from a sky-averaged 21-cm signal experiment requires navigating through an intricate parameter space addressing various factors such as foregrounds, beam uncertainties, ionospheric distortions, and receiver noise for the search of the 21-cm signal. The traditional likelihood-based sampling methods for modeling these effects could become computationally demanding for such complex models, which makes it infeasible to include physically motivated 21-cm signal models in the analysis. Moreover, the inference is driven by the assumed functional form of the likelihood. We demonstrate how Simulation-Based Inference through Truncated Marginal Neural Ratio Estimation ({\small TMNRE}) can naturally handle these issues at a reduced computational cost. We estimate the posterior distribution on our model parameters with {\small TMNRE} for simulated mock observations, incorporating beam-weighted foregrounds, physically motivated 21-cm signal, and radiometric noise. We find that maximizing information content by analyzing data from multiple time slices and antennas significantly improves the parameter constraints and enhances the exploration of the cosmological signal. We discuss the application of {\small TMNRE} for the current configuration of the REACH experiment and demonstrate its potential for exploring new avenues.
\end{abstract}




\section{Introduction}
\import{sections/}{introduction.tex}

\section{Simulation-Based inference}
\label{sec:SBI}
\import{sections/}{mnre.tex}

\section{Simulations and training data}
\label{sec:Sims}
\import{sections/}{simulations.tex}

\section{Results}
\label{sec:results}
\import{sections/}{results.tex}

\section{Summary}
\label{sec:summary}
\import{sections/}{summary.tex}

\section*{Acknowledgements}
We thank the Center for Information Technology of the University of Groningen for their support and for providing access to the Hábrók high performance computing cluster. P.D.M.\ acknowledges support from the Netherlands organization for scientific research (NWO) VIDI grant (dossier 639.042.730). C.W. is supported by the European Research Council (ERC) under the European Union’s Horizon 2020 research and innovation programme (Grant agreement No. 864035 - Undark).  The project has been partially funded by the Netherlands eScience Center, grant number ETEC.2019.018.
 
\section*{Data Availability}
The data underlying this article will be shared on reasonable request to the corresponding author.



\bibliographystyle{mnras}
\bibliography{references} 




\appendix
\section{2D marginal posteriors}
\label{appendix}
\begin{figure*}
    \centering
    \includegraphics[width=\linewidth]{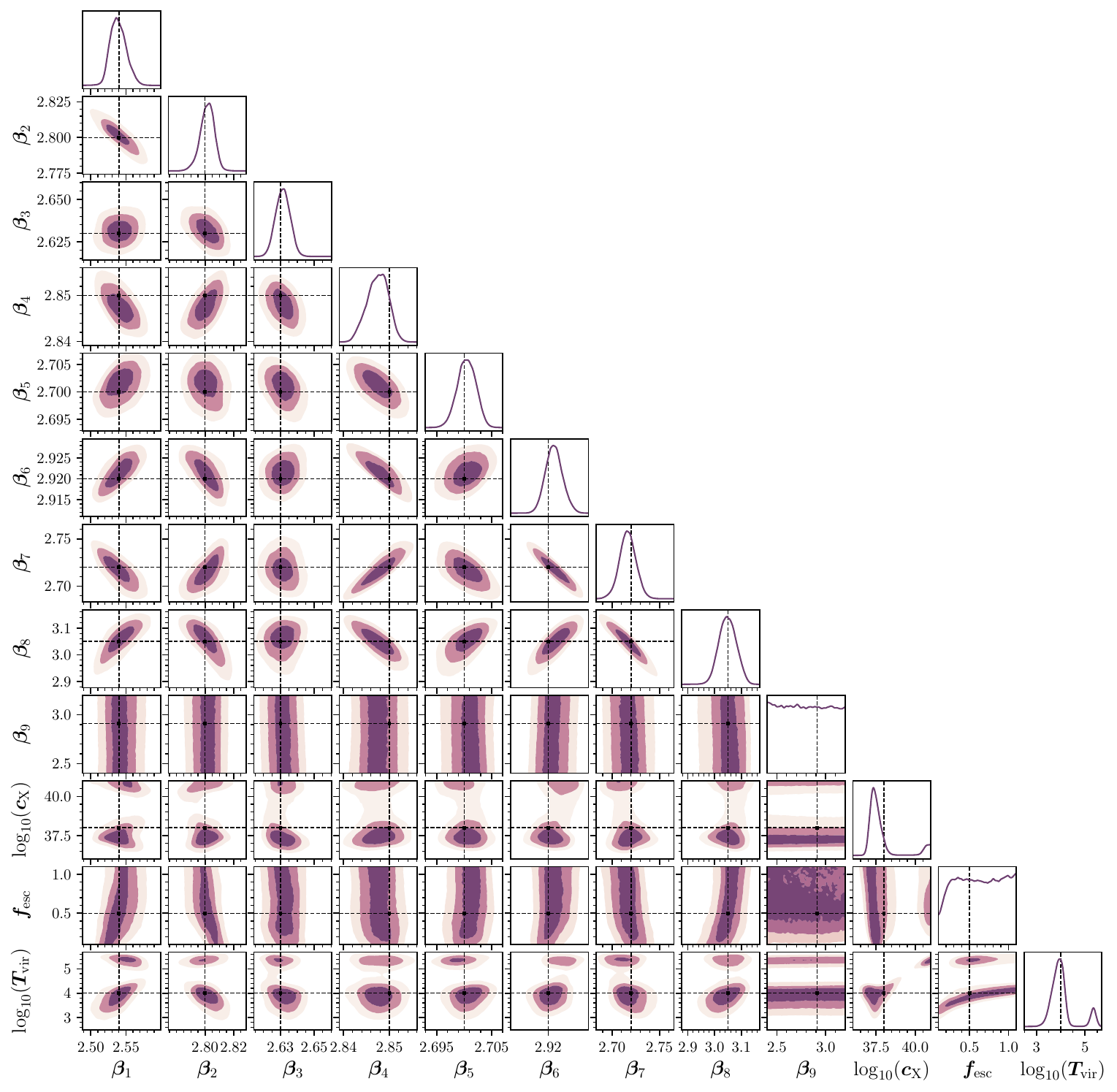}
    \caption{Recovered 1D and 2D marginals on the foreground and the 21-cm signal parameters for the simulated observation for a hexagonal dipole antenna and a single time slice integration. The dashed lines represent the ground truth.}
    \label{fig:post_2d}
\end{figure*}

In this section, we explore the application of {\small TMNRE} to investigate correlations among model parameters through the estimation of 2D marginal posteriors. Specifically, we input pairs of parameters along with the data into a Multi-Layer Perceptron (MLP) to estimate the likelihood-to-evidence ratio for each parameter pair. Figure~\ref{fig:post_2d} presents the 1D and 2D marginal posteriors for the simulated mock observation for a hexagonal dipole antenna and a single time slice ($n_{\rm LST}=1$). The results reveal significant correlations among the foreground parameters i.e.\ spectral indices. Furthermore, several spectral indices are also correlated with the 21-cm signal parameters in our physically motivated model. Notably, the parameters $\log_{10}(c_{\rm X})$ and $\log_{10}(T_{\rm vir})$ exhibit a bi-modal distribution. These parameter degeneracies could be resolved by leveraging the time and beam dependence of the foregrounds, which reduces the correlations between the foregrounds and the 21-cm signal, as shown in Figure~\ref{fig:truncVol_future}.


\bsp	
\label{lastpage}
\end{document}

%% file: sections/introduction.tex
Cosmic Dawn (CD) marks the formation of the first stars and galaxies, which emitted high energy radiation in X-ray and UV bands and ionized the neutral hydrogen that filled the primordial Cosmos during the so-called Epoch of Reionization (EoR) \citep{2001PhR...349..125B,2006PhR...433..181F, Pritchard_2012}. The current astrophysical constraints on CD-EoR are derived from the observations of high redshift quasar spectra \citep{2001AJ....122.2850B, 2003AJ....125.1649F, boera19},  the optical depth of electrons scattering from the Cosmic Microwave Background (CMB) \citep{2003ApJ...583...24K,2011ApJS..192...18K,2020A&A...641A...6P}, and the luminosity function and clustering properties of Lyman-$\alpha$ emitters \citep{Jensen_2012, Dijkstra_2014, Gangolli_2020}. These indirect probes fail to offer a precise understanding of the physical processes unfolding during CD-EoR. The hyperfine transition of the neutral hydrogen, characterized by a restframe wavelength of 21-cm, allows us to directly probe these epochs at low frequencies.

Various experiments are already underway to detect the redshifted 21-cm signal. Interferometric observations from LOFAR \citep{mertens20}, GMRT \citep{2013MNRAS.433..639P}, MWA \citep{barry19,li19}, NenuFAR \citep{mertens2021exploring, Munshi_2024}, HERA \citep{DeBoer_2017, Berkhout_2024} and the upcoming SKA \citep{2015aska.confE...1K,mellema2015hi} characterize the spatial fluctuations in the 21-cm brightness temperature using Fourier statistics. In parallel, there is an alternative approach, which is conceptually simpler but more challenging to calibrate, to detect the sky-averaged 21-cm signal with a wide beam antenna system. These experiments include EDGES \citep{2018Natur.555...67B}, SARAS \citep{2018ApJ...858...54S, https://doi.org/10.48550/arxiv.2112.06778}, PRIZM \citep{https://doi.org/10.48550/arxiv.1806.09531}, SCI-HI \citep{Voytek_2014}, LEDA \citep{Price_2018}, DAPPER \citep{burns2021global}, MIST \citep{monsalve2023mapper}, and REACH \citep{2022NatAs...6..984D}. 

A common challenge for all these experiments is the characterization and isolation of the foregrounds, which are 4-6 orders of magnitude brighter than the cosmological 21-cm signal. The key to separating the foregrounds and the 21-cm signal is to exploit the spectral differences between the two. Foregrounds can be considered spectrally smooth in contrast to the 21-cm signal \citep{1999A&A...345..380S}. However, the beam pattern of the observing instrument carries a spatial and spectral dependence, which corrupts the smoothness of the foregrounds. The resulting beam-weighted foregrounds have a significant co-variance with the 21-cm signal, which makes the extraction of the 21-cm signal quite challenging \citep{Vedantham_2013, Bernardi_2015, 2018ApJ...853..187T, Spinelli_2021, Anstey_2021,10.1093/mnras/stad1047}. In addition, distortions due to ionosphere \citep{Vedantham_2013, Shen_2021}, Radio Frequency Interference (RFI) \citep{leeney2023bayesian, anstey2023enhanced}, and receiver uncertainties \citep{Tauscher_2021, razavighods2023receiver} need to be carefully modeled in the data analysis. To enhance the separation of the 21-cm signal from these systematics, one can leverage drift-scan time information \citep{Tauscher_2020, 10.1093/mnras/stad1047}, measurements of all four Stokes parameters \citep{Nhan_2017, Nhan_2019, Tauscher_2020}, and the dipole of the signal \citep{Slosar_2017, Deshpande_2018, ignatov2023measuringcosmological21cmdipole}.

In 2018, the Experiment to Detect the Global EoR Signature (EDGES) reported the observation of an absorption trough of depth $-500_{-500}^{+200}$ mK centered at 78 $\pm$ 1 MHz \citep{2018Natur.555...67B}. However, the characteristics of the identified signal diverge significantly from the anticipated features outlined in standard astrophysical models \citep{Cohen_2017, Cohen_2020, Reis_2021}. This unexpected discrepancy prompts the consideration of either an exotic explanation to enhance the contrast between the radio background temperature and the spin temperature of the 21-cm signal or an unaccounted systematic effect within the data. Two theoretically plausible scenarios for the former explanation involve the excessive cooling of hydrogen gas due to interactions with dark matter \citep{barkana18, Berlin, Liu19} and the existence of an amplified radio background distinct from the CMB \citep{ewall18,feng18,fialkov19,ewall20,reis20}. Nevertheless, a re-examination of the EDGES data indicates that the presence of unaccounted systematics may distort or obscure the signal \citep{2018Natur.564E..32H, 2019ApJ...880...26S, 2020MNRAS.492...22S, Bevins2020}. Additionally, potential deficiencies in the foreground model used to fit the observations could contribute to the observed trough \citep{Tauscher_2020a}. An independent measurement of the radio sky spectrum by SARAS in the 55-85 MHz band, as reported in \citet{https://doi.org/10.48550/arxiv.2112.06778}, suggests at a confidence level of $\sim 2\sigma$ that the signal detected by EDGES may not be entirely of astrophysical origin.

This work is based in the context of the Radio Experiment for the Analysis of Cosmic Hydrogen (REACH)\footnote{\url{https://www.astro.phy.cam.ac.uk/research/research-projects/reach}}. REACH is a wide band experiment encompassing both the CD and EoR to measure the sky-averaged 21-cm signal in 50-170 MHz frequency range. Deployed in RFI-quiet Karoo radio reserve in South Africa, REACH, in its phase I, is observing the southern hemisphere sky with a hexagonal dipole antenna operating between 50-130 MHz \citep{2022NatAs...6..984D, 2022JAI....1150001C}. Phase II is expected to feature more antennas, including a conical log-spiral antenna operating in 50-170 MHz, and systems sensitive to the polarization of the sky radiation.

In order to achieve a successful and convincing detection, the data must be jointly modeled in a Bayesian framework to account for the 21-cm signal, foregrounds, and instrumental effects, with the goal of identifying and separating these components as much as possible, as they are not inherently orthogonal. Physically motivated models of the sky-averaged 21-cm signal incorporate several astrophysical and cosmological parameters which define the shape of the signal \citep{2009MNRAS.393...32T, Santos_2010, Mesinger_2010, Fialkov_2014}. These models can generate a realization of the signal on the order of seconds to minutes. However, fitting such 21-cm signal models jointly with the foregrounds in a Bayesian framework requires $\sim 10^7$ samples, which is computationally expensive. Therefore, existing data analysis and interpretation methods use the shape of a Gaussian \citep{Bernardi:2016pva, Anstey_2021} or a flattened Gaussian \citep{2018Natur.555...67B} to define the 21-cm signal, which can not describe both the absorption and emission part of the signal. The development of neural network based emulators such as \texttt{globalemu} \citep{Bevins_2021}, {\small 21CMVAE} \citep{Bye_2022}, and {\small 21CMEMU} \citep{2024MNRAS.527.9833B} enable fast and efficient parameter estimation. However, such emulators must be very accurate for an unbiased parameter inference \citep{jones2023validating}. 

Moreover, the statistical analysis with the conventional likelihood-based approach relies on the functional form of the likelihood, which is generally assumed to be Gaussian. This assumption leads to inaccurate posterior inferences if the noise structure is a priori unknown \citep{Scheutwinkel_2023}. These methods sample the full joint posterior distribution of parameters and consequently require more samples (and computation time) to converge for high dimensional parameter spaces \citep{Handley_2015}. In the context of the data analysis of any sky-averaged 21-cm signal experiment, such methods could become a bottleneck because of a large number of degrees of freedom, including the foreground parameters, beam uncertainties, 21-cm signal parameters, ionospheric distortions, and receiver effects.

These issues can be resolved by performing a Simulation-Based Inference ({\small SBI}), where deep learning algorithms are used to estimate the posterior distribution of parameters \citep{Alsing_2019, Cranmer_2020, lueckmann2021benchmarking}. In {\small SBI} methods, the likelihood is implicitly defined through forward simulations. There are a variety of methods to perform {\small SBI}; however, in this work, we apply Truncated Marginal Neural Ratio Estimation ({\small TMNRE}) \citep{https://doi.org/10.5281/zenodo.5043706} algorithm for parameter inference. This approach can directly estimate the marginal posteriors for the parameters of interest instead of sampling the full joint posterior, making it significantly more efficient for exploring high-dimensional parameter spaces. This has already been applied in several astrophysical and cosmological parameter inference applications \citep{Cole_2022, montel2022detection, coogan2022walks, 10.1093/mnras/stad2659, alvey2023simulationbased, alvey2023things,bhardwaj2023peregrine, karchev2023simsims, karchev2024sidereal}.

In this work, we perform {\small SBI} of the foregrounds and 21-cm signal for simulated mock observations for the REACH experiment requiring up to an order of magnitude fewer simulations than the likelihood-based sampling methods. We demonstrate the flexibility of our framework by applying it, for the first time, on a physically motivated 21-cm signal model for the current setup of REACH. We further discuss how this algorithm can be used to explore potential avenues for REACH phase II. This article is organized as follows: In section~\ref{sec:SBI}, we give a brief overview of {\small SBI} and its implementation through {\small TMNRE}. In section~\ref{sec:Sims}, we describe the modeling of the foregrounds and 21-cm signal. In section~\ref{sec:results}, we present the inference of foregrounds and cosmological signal for two different models of the 21-cm signal for the REACH experiment. In section~\ref{sec:summary}, we summarize our findings and discuss the scope for future work.

%% file: sections/mnre.tex
\begin{figure*}
    \centering
    \includegraphics[width=\linewidth]{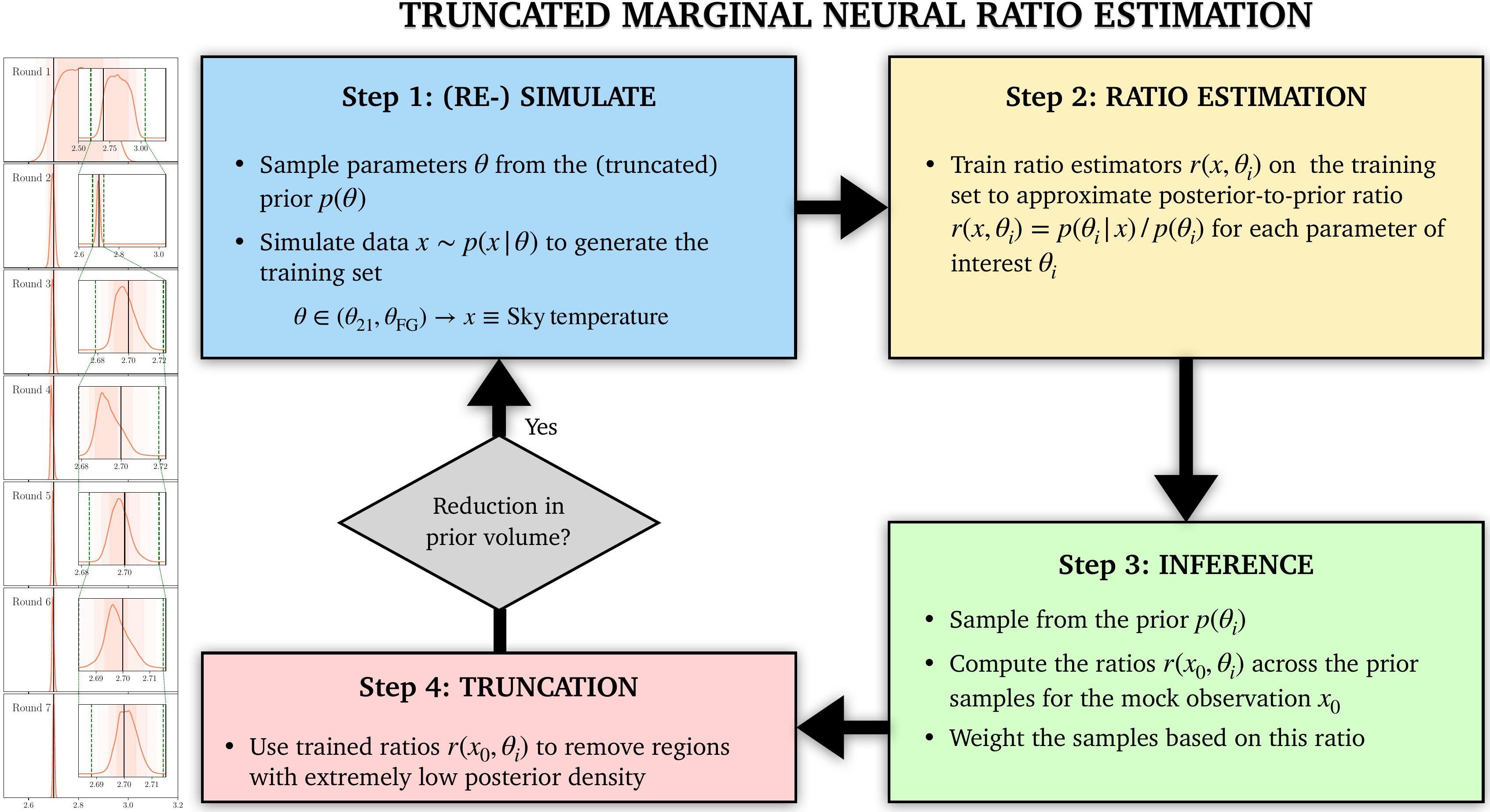}
    \caption{A schematic view of the Truncated Marginal Neural Ratio Estimation ({TMNRE}) algorithm as adopted in our SBI framework.}
    \label{fig:tmnre_algo}
\end{figure*}
In this section, we give a brief overview of Simulation-Based Inference (SBI) \citep{Alsing_2019, Cranmer_2020, lueckmann2021benchmarking}, and its implementation as adopted in this work. In the past four to five years, there has been a notable rise in the advancement and utilization of SBI approaches for analyzing data \citep{Cranmer_2020}. Specifically, the central question guiding various algorithms is \textit{whether a robust Bayesian inference is achievable for a given generative model.}

To be more precise, consider a forward generative model, $p(\*x, \*\theta)$, which takes some underlying set of parameters $\*\theta$ and generates simulated data $\*x$. In a Bayesian sense, the forward model takes the form $p(\*x, \*\theta) = p(\*x|\*\theta)\,p(\*\theta)$, where $\*\theta$ is sampled from a prior distribution $p(\*\theta)$. This expression represents the key functionality of SBI-based methods since running a forward generative model is equivalent to sampling from the simulated data likelihood $p(\*x|\*\theta)$. Scientifically, we are interested in the posterior distribution $p(\*\theta|\*x)$ of the parameters $\*\theta$ for an observation $\*x$, which follows from Bayes' theorem
\begin{equation}
    \label{eq:bayes}
    p(\*\theta|\*x) = \frac{p(\*x|\*\theta)}{p(\*x)}\, p(\*\theta)\,,
\end{equation}
where $p(\*x|\*\theta)$ is the likelihood of the data $\*x$ for given parameters $\*\theta$, $p(\*\theta)$ is the prior probability distribution of the parameters and $p(\*x)$ is the evidence of the data.

Given this setup and the ability to sample from the forward model, there are three main approaches in which SBI methods construct posterior densities. The first approach is \textit{Neutral Posterior Estimation} \citep{papamakarios2018fast, Tejero-Cantero2020, zeghal2022neural, dax2023group}, where the goal is to represent the posterior density $p(\*\theta|\*x)$ with some flexible density estimators such as normalizing flow. The second class of methods falls under the category of \textit{Neural Likelihood Estimation} \citep{papamakarios2018fast, Alsing_2019, Lin_2023}, where the goal is to have an estimator for the simulated data likelihood $p(\*x|\*\theta)$ to be embedded later in traditional inference techniques such as {\small MCMC} or nested sampling. The final class of methods, which is also the focus of this work, is \textit{Neural Ratio Estimation} \citep{https://doi.org/10.5281/zenodo.5043706}, which approximates the likelihood-to-evidence ratio $p(\*x|\*\theta)/p(\*x)$. This is discussed in detail in the following section.

\subsection{Truncated Marginal Neural Ratio Estimation}
In this work, we use a specific algorithm of {\small SBI} known as Truncated Marginal Neural Ratio Estimation ({\small TMNRE}). The two critical features of this method, which make it highly efficient compared to traditional algorithms, are truncation and marginalization. We will first discuss the Neural Ratio Estimation part of this approach and later discuss how these features come into play.

\paragraph*{Neural Ratio Estimation:} In {\small SBI}, the information about the likelihood is implicitly accessed through a stochastic simulator, which defines a mapping from input parameters $\*\theta$ to a sample $\*x \sim p(\*x|\*\theta)$. We sample from this simulator to generate sample-parameter pairs $\{(\*x^1, \*\theta^1)\, (\*x^2, \*\theta^2), \cdots\}$. Here, $\*\theta^i$ is typically drawn from the prior, so the pairs are drawn from the joint distribution $p(\*x, \*\theta)$. These pairs are then employed to train a neural network to approximate the likelihood-to-evidence ratio, a procedure known as Neural Ratio Estimation ({\small NRE}). This ratio can be expressed as
\begin{equation}
    r(\*x, \*\theta) \equiv \frac{p(\*x|\*\theta)}{p(\*x)} = \frac{p(\*\theta|\*x)}{p(\*\theta)} = \frac{p(\*x, \*\theta)}{p(\*x)\,p(\*\theta)}\,.
\end{equation}
In other words, $r(\*x, \*\theta)$ is equal to the ratio of the joint probability density $p(\*x, \*\theta)$ to the product of marginal probability densities $p(\*x)\, p(\*\theta)$. This method then takes the form of a binary classification task to distinguish between jointly-drawn and marginally-drawn pairs. A binary classifier $d_{\*\phi}(\*x, \*\theta)$ with some learnable parameters $\*\phi$ is then optimized to perform this classification. The parameters $\*\phi$ of the model are updated as the model is trained.

For optimizing the model $d_{\*\phi}(\*x, \*\theta)$, we introduce a binary label $y$ to differentiate between the jointly drawn ($y=1$) and marginally drawn ($y=0$) sample-parameter pairs. To be precise, $y$ is a random variable. Once the classifier is trained, it approximates the probability that a sample-parameter pair ($\*x, \*\theta$) is drawn jointly ($y=1$), i.e., \
\begin{equation*}
    \begin{split}
        d_{\*\phi} (\*x, \*\theta) &\approx p(y=1|\*x, \*\theta) \\
        &= \frac{p(\*x, \*\theta|y=1) p(y=1)}{p(\*x, \*\theta|y=1) p(y=1) + p(\*x, \*\theta|y=0) p(y=0)} \\
        &= \frac{p(\*x, \*\theta)}{p(\*x, \*\theta) + p(\*x) p(\*\theta)}\,,
    \end{split}
\end{equation*}
where we assumed $p(y=0) = p(y=1) = 0.5$. This learning problem is associated with a binary cross-entropy loss function
\begin{equation}
    - \int \left[p(\*x, \*\theta) \ln d_\phi (\*x, \*\theta) + p(\*x)\, p(\*\theta) \ln \{1 - d_\phi (\*x, \*\theta)\}\right] \dd{\*x} \dd{\*\theta}\,,
\end{equation}
which is minimized using stochastic gradient descent to find the optimal parameters $\*\phi$ of the network. Once the network is trained, it results in 
\begin{equation}
    d_{\*\phi} (\*x, \*\theta) \approx \frac{p(\*x, \*\theta)}{p(\*x, \*\theta) + p(\*x) p(\*\theta)} = \frac{r(\*x, \*\theta)}{r(\*x, \*\theta) + 1}\,,
\end{equation}
which can be re-written as 
\begin{equation}
        r(\*x, \*\theta) \approx \frac{d_{\*\phi}(\*x, \*\theta)}{d_{\*\phi}(\*x, \*\theta)-1} \implies
        p(\*\theta|\*x) \approx \frac{d_{\*\phi}(\*x, \*\theta)}{d_{\*\phi}(\*x, \*\theta)-1}\,p(\*\theta)
\end{equation}
to estimate the posterior probability distribution. In practice, this classifier is simply a dense neural network with a few hidden layers.

The two key features of this method are the following.
\begin{itemize}
    \item It can directly estimate marginal posteriors by omitting model parameters from the network's input instead of first sampling the full joint posterior and then performing the marginalization and,
    \item The analysis is done sequentially in multiple rounds. The posterior densities are estimated at the end of each round for the given observation. The prior for the next round is then truncated based on the inference in the current round, hence the name Truncated Marginal ({\small NRE}). This leads to a targeted inference where extremely low posterior density regions are discarded after each round.
\end{itemize}

\begin{table}
 \centering
 \caption{Settings for the TMNRE algorithm used in this work.}
 \label{tab:TMNRE}
 \def\arraystretch{1.}
 \begin{tabular}{ll}
  \hline
  Setting & Value\\
  \hline
  \hline
  Initial learning rate & 5 $\times 10^{-4}$\\
  Max training epochs & 70\\
  Early stopping & 20\\
  Number of rounds & 7\\
  Simulation schedule & 30k, 30k, 60k, 90k, 150k, 300k, 300k\\
  Bounds threshold & $10^{-3}$\\
  \hline
 \end{tabular}
\end{table}
Both these features significantly reduce the simulation budget compared to likelihood-based sampling methods. In this work, we use {\small TMNRE} as implemented in the software package \texttt{swyft}\footnote{\url{https://github.com/undark-lab/swyft}}. This algorithm is summarized in Figure~\ref{fig:tmnre_algo}. The four basic steps involved in this framework are as follows.
\begin{enumerate}
    \item \textbf{(Re-)Simulate:} Generate samples $\*x \sim p(\*x|\*\theta)$ based on the current prior distribution $p(\*\theta)$ of model parameters\footnote{This step is fully parallelized in our implementation which makes it significantly fast with access to appropriate hardware.}.\vspace{0.5em}
    \item \textbf{Ratio Estimation:} Train ratio estimators $r(\*x, \*\theta)$ for each parameter of interest $\theta_i$ using samples from step (i) to approximate the posterior-to-prior ratio $p(\theta_i|\*x)/p(\theta_i)$. \vspace{0.5em}
    \item \textbf{Inference:} For a given mock observation $x_0$, estimate the ratios $r(x_0, \theta_i)$ across the prior samples $p(\theta_i)$, and weight the samples based on this ratio. \vspace{0.5em}
    \item \textbf{Truncation:} Apply truncation by removing regions with extremely low posterior density based on the weighted samples from step (iii). If the prior volume is reduced, then re-simulate from step (i) with the truncated prior.
\end{enumerate}

Our {\small TMNRE} setup is implemented with the settings listed in Table~\ref{tab:TMNRE}. These include the initial learning rate, the maximum number of epochs to train the network, the number of epochs to wait for the validation loss to increase (early stopping), the number of rounds to train the network, simulation schedule i.e.\ the number of samples to be used in each round and bounds threshold for applying truncation.

%% file: sections/simulations.tex
\begin{figure*}
    \centering
    \includegraphics[width=0.32\linewidth]{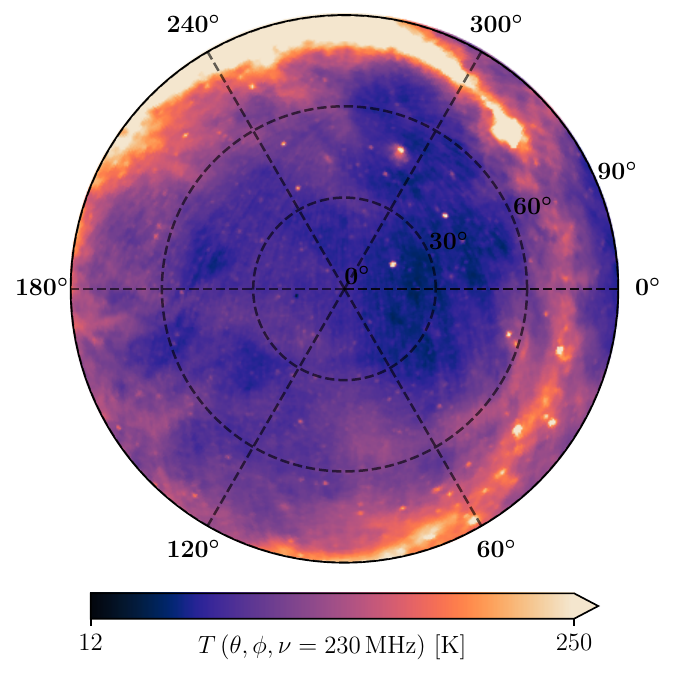}\hspace{0.5em}
    \includegraphics[width=0.32\linewidth]{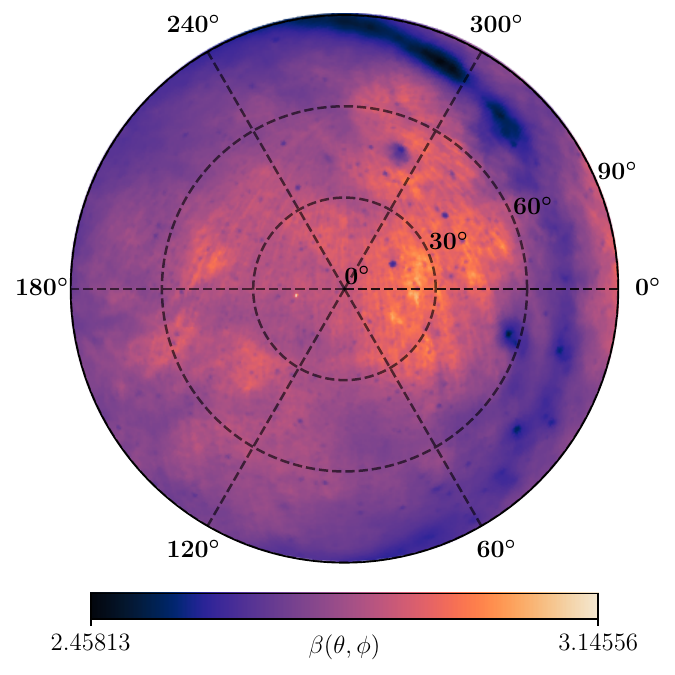}\hspace{0.5em}
    \includegraphics[width=0.32\linewidth]{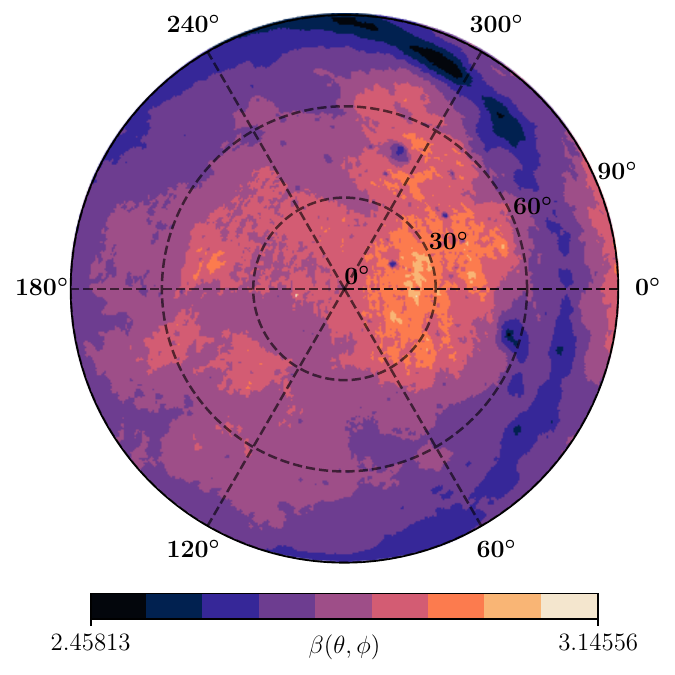}
    \caption{Radio sky GSM at 230 MHz (\textit{left panel}), spectral index distribution (\textit{middle panel}), and the division of the sky into nine regions (\textit{right panel}). These projections correspond to a simulated observation on 2019-10-01 at 00:00:00 UTC when the galactic disk remains below the horizon ($\theta > 90^{\circ}$) for an antenna located in the Karoo radio reserve. These are shown in co-latitude ($\theta$) and azimuth ($\phi$) coordinates of the antenna's frame with the zenith in the center.}
    \label{fig:gsm_map}
\end{figure*}  

This section describes our forward modeling of the foregrounds and the 21-cm signal. We assume an observation ($\*T_{\rm obs}$) to be composed of the sky-averaged 21-cm signal ($\*T_{21}$), beam-weighted foregrounds ($\*T_{\rm FG}$), and a radiometric noise component given as $\*\sigma = (\*T_{21} + \*T_{\rm FG})/\sqrt{\Delta \nu \Delta t}$, where $\Delta \nu$ is the channel width and $\Delta t$ is the integration time. The forward model then takes the form
\begin{equation}
    \*T_{\rm obs} (\nu) = \*T_{\rm 21} (\nu) + \*T_{\rm FG} (\nu) + \*{\sigma} (\nu)\,.
\end{equation}

There are also other sources of error in the data due to imperfect calibration of the instrument, additive and multiplicative biases through the receiver's gain. In this work, we assume that the instrument is well-calibrated and do not take into account these sources of uncertainty. These effects will be considered in a future work.

\subsection{Modeling the foregrounds}
A beam-weighted foreground model can be constructed from the following features: (i) a spatial brightness temperature distribution, (ii) a spatial distribution of spectral index, and (iii) the spatial and spectral dependence of the observing antenna. 

We generate our sky model as
\begin{equation}
    T_{\rm sky} (\theta, \phi, \nu) = \left[ T_{\rm 230} (\theta, \phi) - T_{\rm CMB}\right] \left( \frac{\nu}{230} \right)^{-\beta(\theta, \phi)} + T_{\rm CMB}\,,
\end{equation}
where $T_{230} (\theta, \phi)$ is the spatial brightness temperature distribution derived from the Global Sky Map (GSM) at 230 MHz \citep{de_Oliveira_Costa_2008}, which we use as our base foreground map, and $T_{\rm CMB} = 2.725$ K. Our simulated observations start at 2019-10-01 00:00:00 UTC and are integrated for 6 hours. In the left panel of Figure~\ref{fig:gsm_map}, we show the overhead sky at 230 MHz on 2019-10-01 00:00:00 UTC for an antenna located in the Karoo radio reserve in South Africa. 

\begin{figure*}
    \centering
    \includegraphics[width=0.97\linewidth]{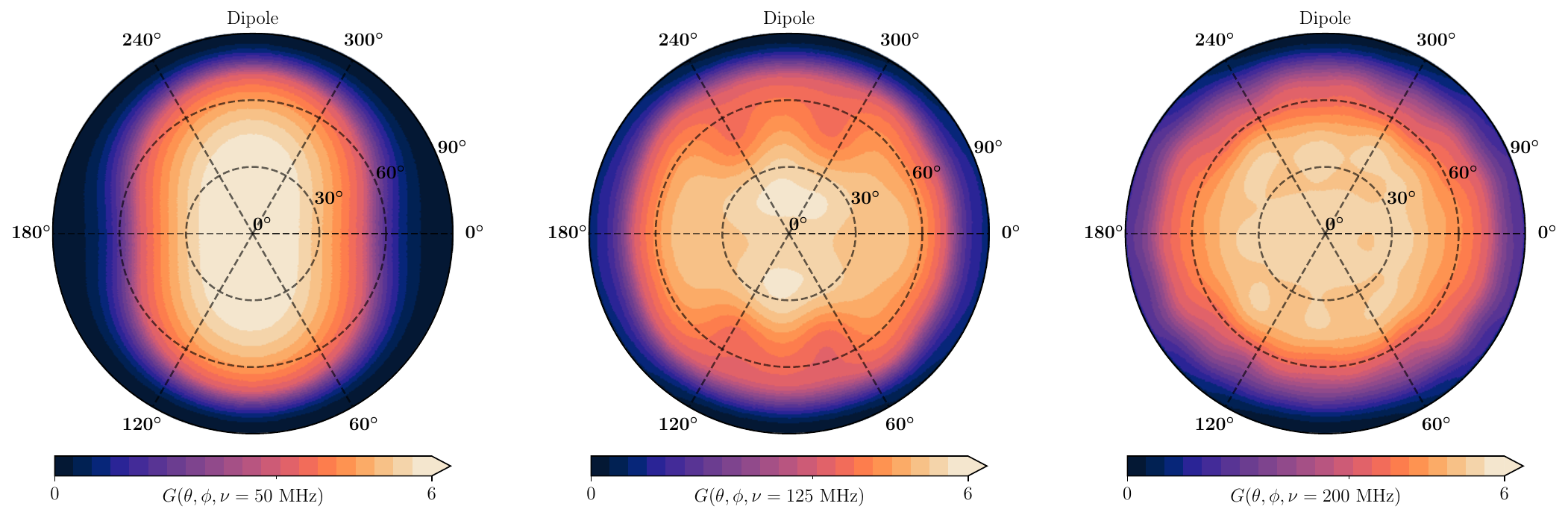} 
    \includegraphics[width=0.97\linewidth]{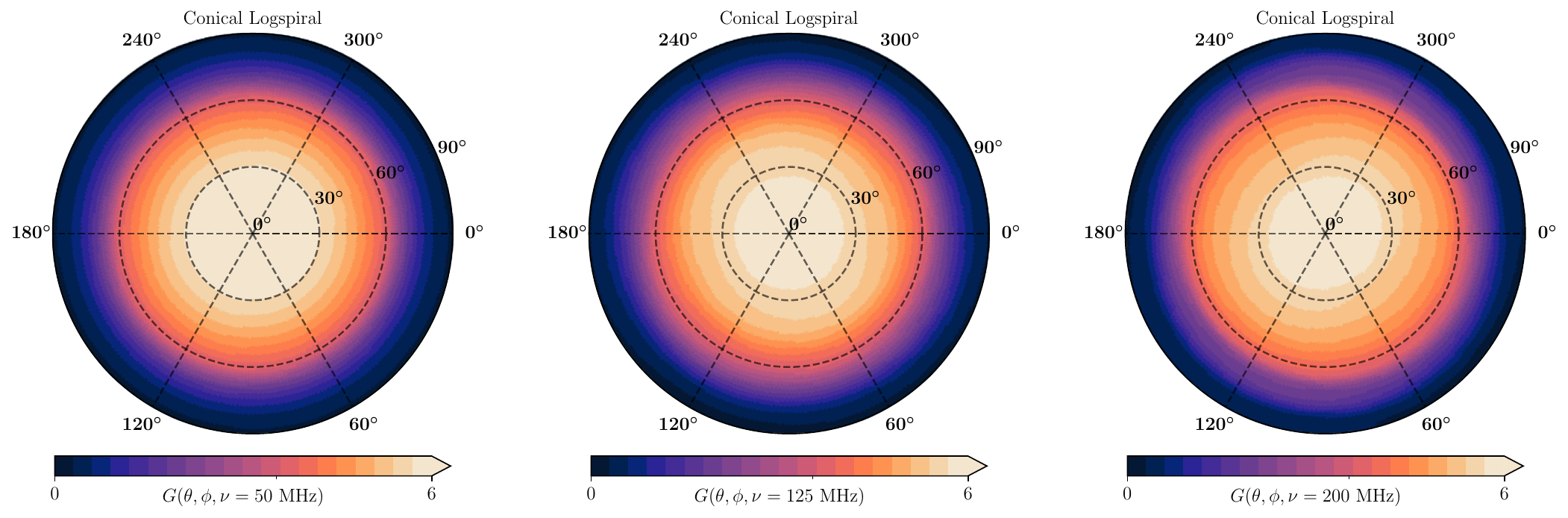}
    \caption{Beam patterns $G(\theta, \phi, \nu)$ for the dipole (top row) and conical log spiral (bottom row) antenna at $\nu$ = 50 MHz (first column), 125 MHz (second column) and 200 MHz (third column) shown in the co-latitude ($\theta$) and azimuth ($\phi$) coordinates of the antenna's frame.}
    \label{fig:beam_patterns}
\end{figure*}

Next, for deriving a spectral index variation map, we apply a pixel-wise tracing between an instance of GSM at 230 MHz and 408 MHz as
\begin{equation}
    \beta(\theta, \phi) = \frac{\log{\left(\frac{T_{230} (\theta, \phi) - T_{\rm CMB}}{T_{408} (\theta, \phi) - T_{\rm CMB}}\right)}}{\log{\left(\frac{230}{408}\right)}}\,.
\end{equation}

The resulting spectral index map is shown in the middle panel of Figure~\ref{fig:gsm_map}. To model the foregrounds in a parameterized manner, we divide the sky into nine regions and assign a constant spectral index to each region. This physically motivated modeling of the foregrounds follows from \citet{Anstey_2021}. Our choice of using nine regions is based on the findings of \citet{Anstey_2021} for the simulated observations of REACH. However, the number of regions required to adequately model the data also depends on the number of time slices and the chromaticity of the antenna beam. Therefore, the selection of the number of regions should be data-driven, based on Bayesian evidence \citep{Anstey_2021, 10.1093/mnras/stad156, 10.1093/mnras/stad3392}. For modern SBI algorithms, the recently proposed learned harmonic mean estimator \citep{mcewen2023machinelearningassistedbayesian} offers a promising method for estimating Bayesian evidence \citep{Spurio_Mancini_2023}, which we plan to investigate in future work. These regions are formed by dividing the full range of spectral indices as shown in the middle panel of Figure~\ref{fig:gsm_map} into nine equal intervals, and each region is defined as the patch of the sky with spectral indices within these intervals. The parameterized foreground model $T_{\rm sky}(\theta, \phi, \nu)$ then takes the form
\begin{equation}
    \label{eq:tsky_param}
    \left[\sum_{i=1}^{9} M_i(\theta, \phi) \left(T_{\rm 230} (\theta, \phi) - T_{\rm CMB}\right) \left(\frac{\nu}{230}\right)^{-\beta_i}\right] + T_{\rm CMB}\,,
\end{equation}
where $M_i (\theta, \phi)$ represents a masks for the sky region $i$, which has a spectral index $\beta_i$. $M_i(\theta, \phi) = 1$ for every pixel within the region and $0$ elsewhere. The division of the sky is shown in the right panel of Figure~\ref{fig:gsm_map}, where the spectral index for each region is randomly sampled from $\beta_i \in (2.4, 3.2)$ for $1 \leq i \leq 9$.

This parameterized sky model is convolved with the beam of the antenna to form beam-weighted foregrounds and then averaged over the sky
\begin{equation}
    \label{eq:bwfg}
    \*T_{\rm FG} (\nu) = \frac{1}{4\upi}\int_{0}^{4\upi} G(\theta, \phi, \nu)\, T_{\rm sky} (\theta, \phi, \nu) \dd{\Omega}\,,
\end{equation}
where $G(\theta, \phi, \nu)$ describes the spatial and spectral dependence of the antenna beam. In this work, we consider two different antenna designs: a hexagonal dipole and a conical log-spiral antenna. Our simulations are run in 50-200 MHz frequency range. The beam patterns of these antennas are shown in Figure~\ref{fig:beam_patterns} at three different frequencies $\nu = 50, 125$, and 200 MHz in the antenna's frame of reference, where the green line marks the horizon.

\begin{figure*}
    \centering
    \includegraphics[width=\linewidth]{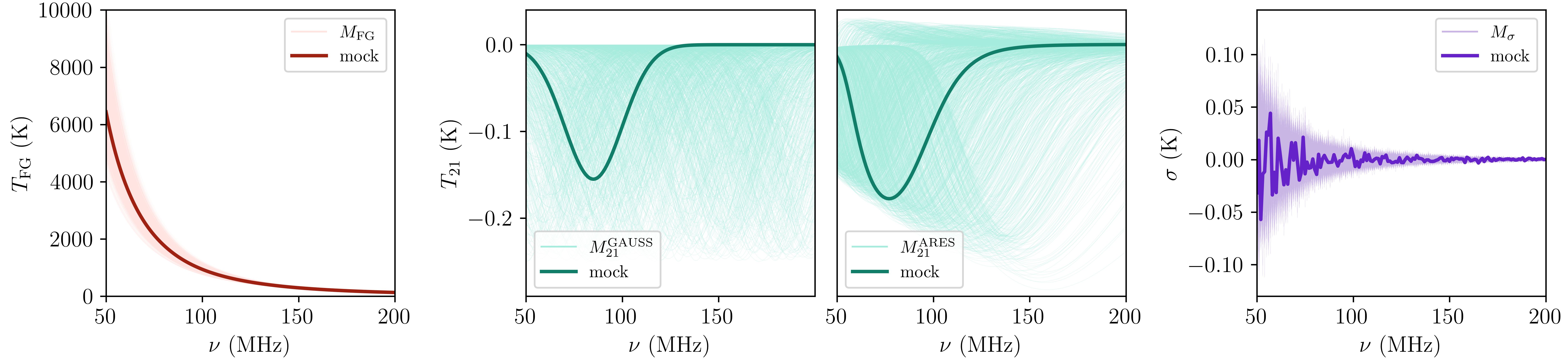}
    \caption{\textbf{Left Panel:} The beam-weighted foreground modeling set ($\*M_{\rm FG}$) for a dipole antenna. \textbf{Center Panel:} The sky-averaged 21-cm signal training set for a Gaussian profile (left) and a physical model (right). \textbf{Right Panel:} Radiometric noise realizations ($\*M_{\rm \sigma}$) for each component of the training set. The dark-colored lines in all panels represent the realization used in the mock observation.}
    \label{fig:mockObs}
\end{figure*}

To generate a realization of the beam-weighted foregrounds, we sample the spectral index for each region $\beta_i \in (2.4, 3.2)$ for $1 \leq i \leq 9$, and use equation~(\ref{eq:tsky_param}) and~(\ref{eq:bwfg}). In the left panel of Figure~\ref{fig:mockObs}, we show a thin slice of the beam-weighted foreground modeling set ($\*M_{\rm FG}$) for a dipole antenna, generated by uniformly sampling over the prior on $\beta_i$. The dark red line represents a random realization used in the mock observation ($\*T_{\rm FG}^{\rm mock}$) with $\beta = \{2.54, 2.8, 2.63, 2.85, 2.7, 2.92, 2.72, 3.05, 2.91\}$.

\subsection{Modeling the 21-cm signal}
In this work, we consider two different types of models (i) a physically motivated model of the sky-averaged 21-cm signal, and (ii) a Gaussian profile, which is generally used as an approximation of the cosmological signal \citep{Anstey_2021, 10.1093/mnras/stad156}. The details of these two models are discussed in the following subsections.

\subsubsection*{Gaussian 21-cm signal model}
For simplicity, our first model of the sky-averaged 21-cm signal takes the form of a Gaussian, which is described as
\begin{equation}
    \*T_{21}(\nu) = -A\, {\rm exp} \left[ \frac{(\nu - \nu_0)^2}{2\sigma^2} \right]\,,
\end{equation}
where $A$ is the amplitude of the absorption feature, $\nu_0$ is the central frequency, and $\sigma$ is the standard deviation. We adopt uniform priors on each of these parameters as follows: $A \in (0\, {\rm K}, 0.25\, {\rm K})$, $\nu_0\, \in (50\, {\rm MHz}, 200\, {\rm MHz})$, and $\sigma \in (10\, {\rm MHz}, 20\, {\rm MHz})$.

In the center-left panel of Figure~\ref{fig:mockObs}, we show a thin slice of the 21-cm signal modeling set ($\*M_{21}^{\rm GAUSS}$) generated with the aforementioned priors. The dark green line represents the realization used in the mock observation ($\*T_{\rm 21}^{\rm mock}$) with $A=0.155\,{\rm K}$, $\nu_0=85\,{\rm MHz}$, and $\sigma=15\,{\rm MHz}$.

\subsubsection*{Physical 21-cm signal model}
To simulate the physically motivated sky-averaged 21-cm signal, we use Accelerated Reionization Era Simulations (\texttt{ARES}\footnote{\url{https://ares.readthedocs.io/en/ares-dev/}}), which is a semi-analytical code based on 1D radiative transfer \citep{2012ApJ...756...94M, 2014MNRAS.443.1211M, 2017MNRAS.464.1365M}. Most sky-averaged 21-cm signal models are semi-numerical and can generate a realization of the 21-cm signal within minutes to hours \citep{Mesinger_2010, Fialkov_2014, Ghara_2015, Murray2020, Schaeffer_2023, Hutter_2023}; however, this is computationally infeasible for an analysis that requires millions of samples. Conversely, the semi-analytical approach of \texttt{ARES} produces a realization of the 21-cm signal within a few seconds. Its efficiency stems from evolving the mean radiation background directly, rather than averaging over large cosmological volumes. For a detailed description of how \texttt{ARES} models the 21-cm signal, we refer the interested reader to Section 2 of \citet{2017MNRAS.464.1365M}.

In our ARES model, we consider three astrophysical parameters that define the shape and amplitude of the cosmological signal: (i) $c_{\rm X}$, which is the normalization factor of X-ray luminosity - SFR relation and governs the efficiency of galaxies in producing X-ray photons, (ii) $f_{\rm esc}$, which is the escape fraction of ionizing photons, and (iii) $T_{\rm vir}$, the minimum virial temperature of a collapsed halo to form star-forming galaxies. We adopt a log-uniform prior on $c_{\rm X} \in (10^{36}, 10^{41})\, {\rm erg}\, {\rm s}^{-1} (M_{\odot}\, {\rm yr}^{-1})^{-1}$, a uniform prior on $f_{\rm esc} \in (0.1, 1)$, and a log-uniform prior on $T_{\rm vir} \in (3\times 10^2\,{\rm K}, 5\times 10^5\,{\rm K})$. The 21-cm signal modeling set $(\*M_{\rm 21}^{\rm ARES})$ generated with these priors is shown in the center-right panel of Figure~\ref{fig:mockObs}. The dark green line shows the realization used in mock observation with $c_{\rm X} = 10^{38}\, {\rm erg}\, {\rm s}^{-1} (M_{\odot}\, {\rm yr}^{-1})^{-1}$, $f_{\rm esc} = 0.5$, and $T_{\rm vir} = 10^4\, {\rm K}$.

\begin{figure*}
    \centering
    \includegraphics[width=\linewidth]{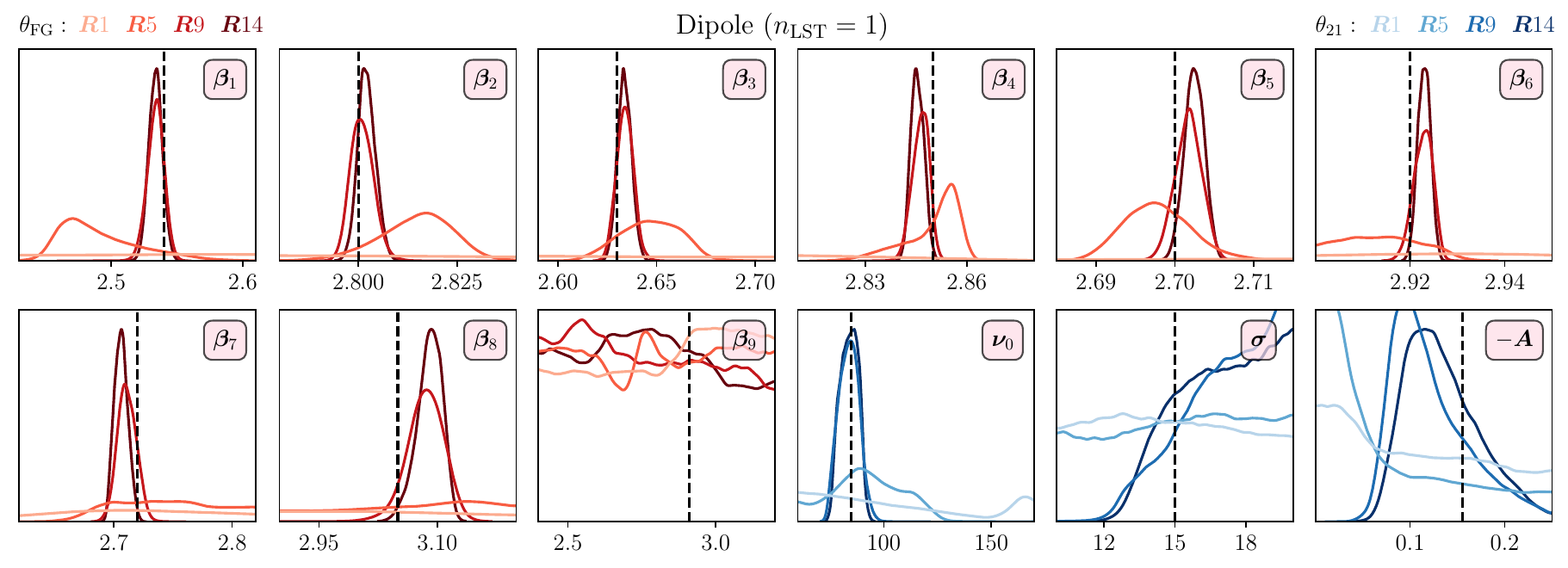}\vspace{2em}
    \includegraphics[height=0.3\linewidth]{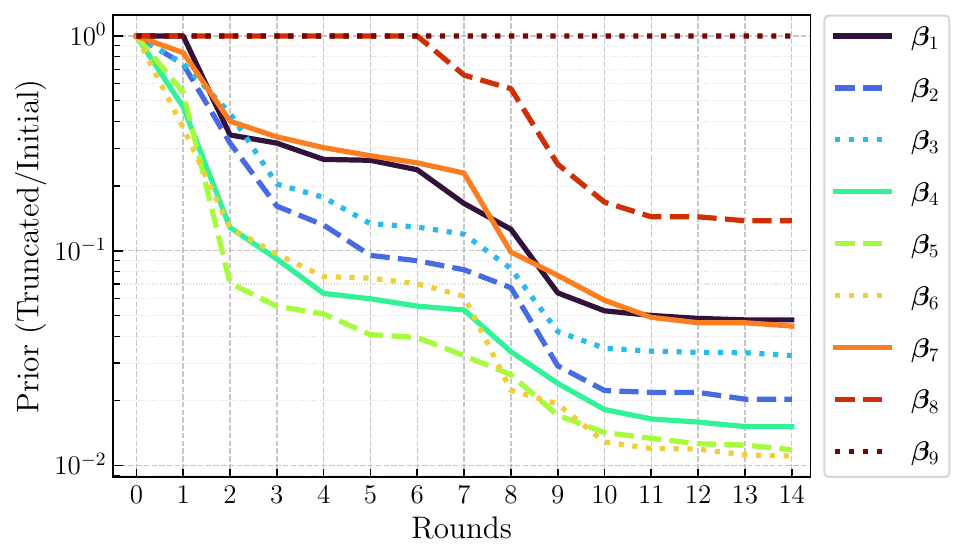}\hspace{1em}
    \includegraphics[height=0.3\linewidth]{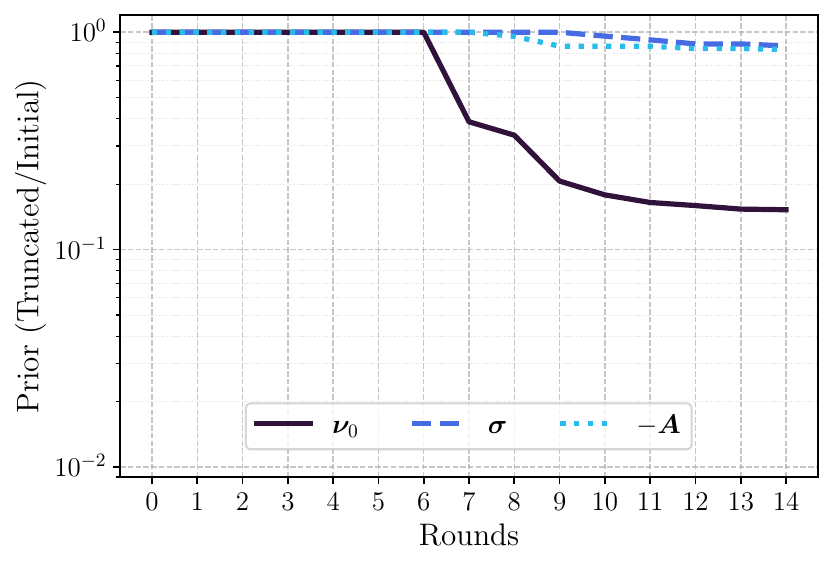}
    \caption{Application of {TMNRE} on the simulated mock observation for a dipole antenna. \textbf{Top panel:} 1D marginal posterior distribution of foreground (red) and 21-cm signal parameters (blue) evaluated after rounds 1, 5, 9, and 14 (shown by different shades). The dashed line represents the true values of the parameters. \textbf{Bottom panel:} The evolution of truncated prior to the initial prior ratio of the foreground (left) and 21-cm parameters (right) as the network learns over multiple rounds.}
    \label{fig:postRounds}
\end{figure*}


%% file: sections/results.tex
\begin{figure*}
    \centering
    \includegraphics[width=\linewidth]{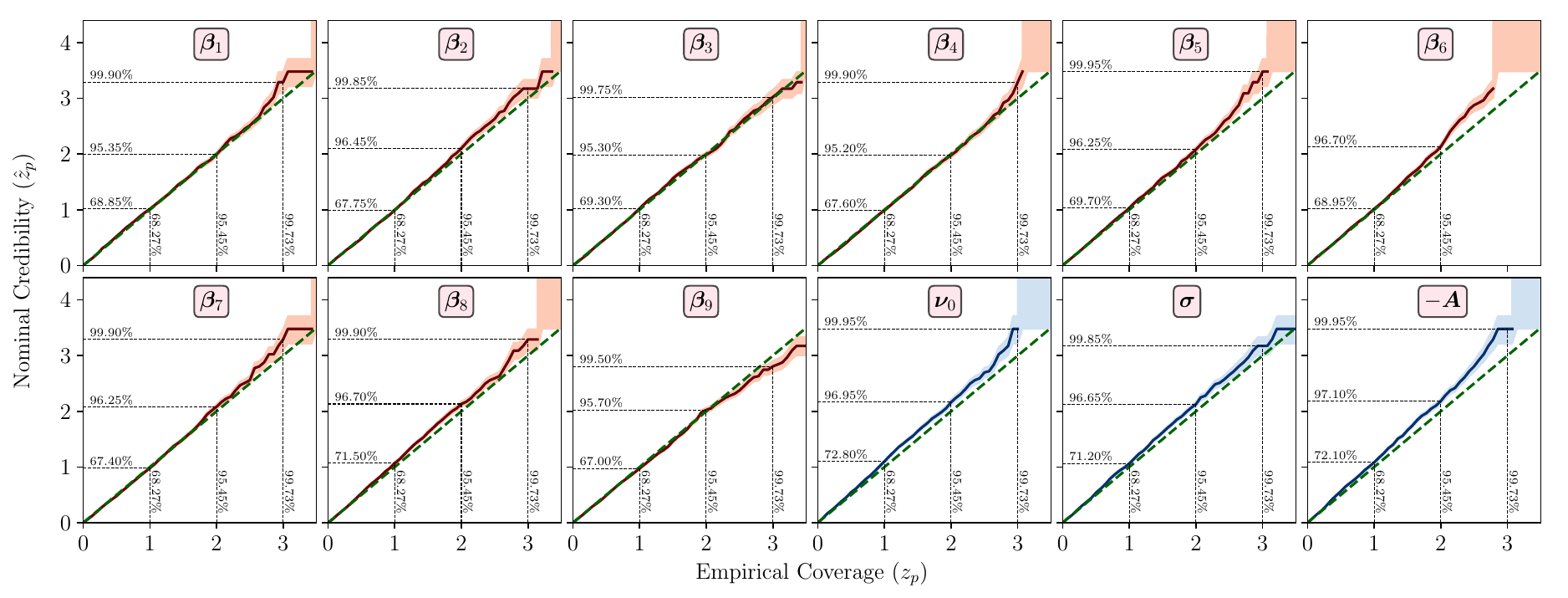}
    \caption{Nominal credibility of the converged network from Figure~\ref{fig:postRounds} as a function of empirical coverage for all foreground (red) and 21-cm signal parameters (blue). The green dashed line represents a perfect coverage.}
    \label{fig:coverage}
\end{figure*}

In this section, we discuss the application of the {\small TMNRE} algorithm to simultaneously constrain the beam-weighted foregrounds and 21-cm signal for our simulated mock observations. We first delve into its application for the current setup of the REACH experiment in section~\ref{sec:REACH_current} and subsequently explore potential avenues to further improve the constraints in section~\ref{sec:REACH_future}.

\subsection{The existing setup of REACH}
\label{sec:REACH_current}
REACH, in its current configuration, is observing the sky with a hexagonal dipole antenna located in the Karoo Radio Reserve in South Africa. In this section, we explore the robustness of this setup to simultaneously constrain the foregrounds and 21-cm signal from the mock data.
\subsubsection*{Gaussian 21-cm signal model}
We first train our neural ratio estimator, assuming a Gaussian profile for the 21-cm signal. In Figure~\ref{fig:postRounds}, we show the truncation of prior for the foreground and 21-cm signal parameters. The top panel shows the 1D marginal posteriors after rounds 1, 5, 9, and 14 (shown by different shades from light to dark). This figure demonstrates how the {\small TMNRE} algorithm performs a targeted inference through truncation. After the first round of training, only a few foreground parameters prior distributions get truncated significantly, for example, $\beta_4$, $\beta_5$, and $\beta_6$. This is expected because the sky regions corresponding to these spectral indices cover the most significant portion of the sky (25.5\%, 31.2\%, 20.5\% respectively) and substantially impact the observed sky-averaged brightness temperature. However, after this truncation, the network is able to learn other parameters in the following rounds. This can be seen in the bottom panel of Figure~\ref{fig:postRounds}, where we show the truncated prior to initial prior ratio for each foreground (left) and 21-cm signal parameter (right).
\begin{figure*}
    \centering
    \includegraphics[width=\linewidth]{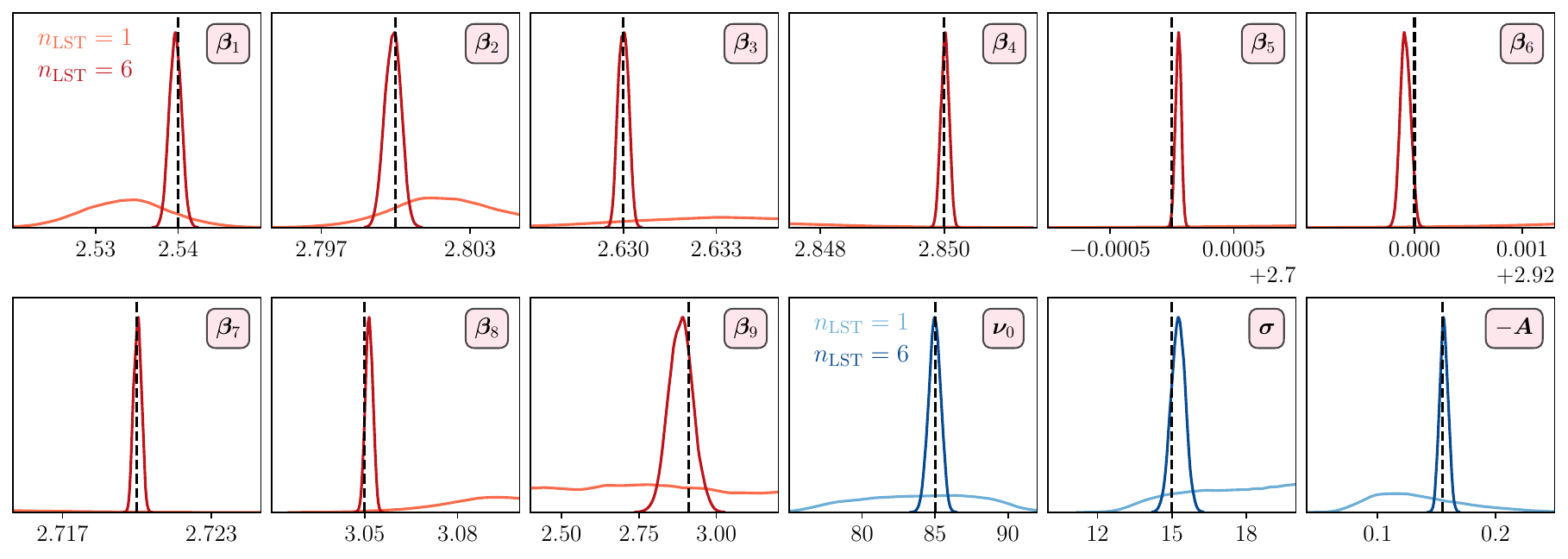}
    \caption{Utilization of drift scan spectra: 1D marginal posteriors of foreground (red) and 21-cm signal parameters (blue) for a dipole antenna when the data is analyzed from a single time slice of 12 hours ($n_{\rm LST}$=1: light) and six time slices of 2 hours each ($n_{\rm LST}$=6: dark). These posteriors are generated once the algorithm is converged for each case. The dashed lines represent the ground truth.}
    \label{fig:post_Gauss_timeBins}
\end{figure*}
\begin{figure*}
    \centering
    \includegraphics[width=\linewidth]{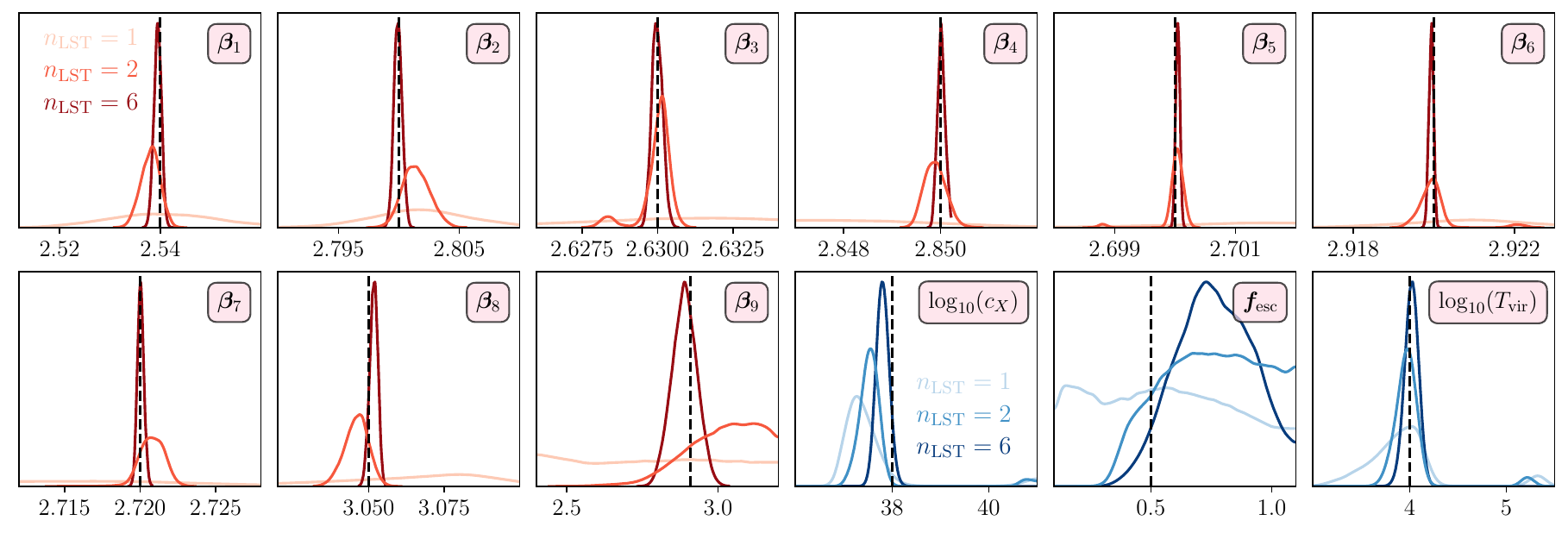}
    \caption{Application of TMNRE to a physically motivated model of the 21-cm signal: 1D marginal posteriors of foreground (red) and 21-cm signal parameters (blue) for a dipole antenna when the data is analyzed using $n_{\rm LST} = 1, 2,$ and 6 (from light to dark). The total integration time is set at 12 hours, which is then evenly divided across multiple time slices. These posteriors are generated once the algorithm is converged for each case.}
    \label{fig:postComp_ARES}
\end{figure*}


The 21-cm signal parameters are learned by the network after the proposal distributions of all foreground parameters undergo a significant truncation. After seven rounds of training, the network is able to truncate the proposal distribution of the central frequency $\nu_0$ of the 21-cm signal. The algorithm converges once we see no further truncation of priors in the following rounds. After fourteen rounds of training, the network can correctly identify the central frequency $\nu_0$ of the cosmological signal along with relatively weak constraints on the absorption amplitude $-A$. However, the width of the signal $\sigma$ remains unconstrained. Once the algorithm converges, the ratio of truncated prior volume to initial prior volume becomes $\sim 10^{-14}$.

In order to validate our inference results, one way is to compare them with traditional sampling-based methods, which are extremely costly. Therefore, it becomes crucial to have additional methods to cross-check and validate our inference \citep{hermans2022trust, lemos2023samplingbased}. In {\small SBI}, this is commonly achieved by estimating the statistical coverage of the trained network. We test the statistical properties of our inference by quantifying the nominal and empirical expected coverage probabilities. The reasoning behind this test is that given a set of $n$ independent and identically distributed samples $(\*x_i, \*\theta_i^*) \sim p(\*x, \*\theta)$, the $x\%$ credible interval for the posterior should contain the ground truth $\*\theta_i^*$ in $x\%$ of the $n$ samples for a well-calibrated posterior distribution. In practice, once the algorithm is converged, we take a large number of samples $n=2000$ from the truncated prior and perform inference on each sample. The local amortization in \texttt{swyft} \citep{https://doi.org/10.5281/zenodo.5043706} allows us to quickly generate posteriors for any given sample. In Figure~\ref{fig:coverage}, we show the nominal credibility as a function of the empirical coverage for all model parameters. The uncertainty on the nominal credibility follows from the finite number of samples ($n$) and is estimated using the Jeffreys interval \citep{Cole_2022}. The green dashed line indicates a perfect coverage. Across all parameters, we find an excellent posterior coverage.

In the analysis of the data from a single time slice ($n_{\rm LST}$=1), as shown in Figure~\ref{fig:postRounds}, the constraints on the 21-cm signal parameters are weak because of a significant co-variance between the beam-weighted foregrounds and the cosmological 21-cm signal. The overlap between these two components can be reduced by utilizing the time dependence of the foregrounds \citep{Liu_2013, 2014ApJ...793..102S, Tauscher_2020, 10.1093/mnras/stad156, 10.1093/mnras/stad1047}. This can be achieved by fitting the data from multiple time slices, and utilizing the fact that the foregrounds change from one time slice to another, but the 21-cm signal appears identically in each time slice. For this purpose, we simultaneously fit the data $\*T_{\rm obs}(\nu)$ from six time slices by giving it as the input to the multi-layer perceptron.

In Figure~\ref{fig:post_Gauss_timeBins}, we compare the posterior distribution of the model parameters from a single time slice (light) and six time slices (dark). We note that the integration time is 12 hours for both scenarios. We find that simultaneously analyzing multiple time slices improves the parameter constraints significantly. By utilizing the time dependence of the foregrounds, the network is able to constrain $\beta_9$, which is relatively hard to infer since the region corresponding to this parameter covers a tiny portion (0.04$\%$) of the sky. Also, the algorithm can now correctly identify the width ($\sigma$) of the 21-cm signal. In this case, the ratio of truncated prior to initial prior volume goes down to $\sim 10^{-26}$ compared to $\sim 10^{-14}$ for a single time slice. Moreover, by increasing the information content, the algorithm converges after seven rounds of training, which is computationally more efficient than the former case.

\subsubsection*{Physical 21-cm signal model}
Next, we focus on a physically motivated model of the sky-averaged 21-cm signal, \texttt{ARES}. As discussed earlier, our 21-cm signal model is composed of three astrophysical parameters $(c_{\rm X}, f_{\rm esc}, T_{\rm vir})$, which is then used in the simulated data analysis. Such models are computationally quite expensive to be explored in the traditional likelihood-based methods because the fitting procedure requires millions of samples. However, this problem can be circumvented within an SBI-based framework, thanks to the efficiency of such algorithms. To increase the efficiency of our algorithm in this case, we first generate training samples of the 21-cm signal from the initial prior range.  As we perform the {\small TMNRE}, we do not apply truncation on the 21-cm signal parameters during the initial rounds of training until all the foreground parameters undergo a significant truncation. In this way, we re-use the samples of 21-cm signal during the initial rounds of training. Once the foreground parameters begin to converge, we apply the truncation on 21-cm signal parameters. This approach used 500,000 21-cm signal samples to estimate the posteriors with the data from a single time slice, which is an order of magnitude fewer compared to a likelihood-based framework \citep{Anstey_2021}. 

In Figure~\ref{fig:postComp_ARES}, we present the posterior distribution of our model parameters for the \texttt{ARES} 21-cm signal model when the data is analyzed for $n_{\rm LST}=1, 2$, and 6 (from light to dark). Applying {\small TMNRE} on the data from a single time slice, the network is able to constrain the efficiency of X-ray photon production $c_{\rm X}$ and minimum virial temperature of a halo to host star-forming galaxies $T_{\rm vir}$, with no constraints on the escape fraction of ionizing photons $f_{\rm esc}$. The constraints on $c_{\rm X}$ and $T_{\rm vir}$ are further improved with the inclusion of multiple time slices with relatively weak constraints on $f_{\rm esc}$. This is expected because the escape fraction of UV photons $f_{\rm esc}$ weakly affects the emission part of the cosmological signal. Moreover, for $n_{\rm LST} = 1$ and 2, the posteriors of $c_{\rm X}$ and $T_{\rm vir}$ are bi-modal due to degeneracy between the foreground and 21-cm signal parameters (as shown in Appendix~\ref{appendix}). This degeneracy is mitigated with the addition of more time slices, particularly when $n_{\rm LST} = 6$.

\subsection{Future possibilities}
\label{sec:REACH_future}
In this section, we discuss potential avenues for the REACH experiment. As mentioned in \citet{2022NatAs...6..984D}, REACH in Phase II will feature more antennas to observe the southern hemisphere sky for a better understanding and isolation of the signal components associated with hardware systematics. The follow-up configurations might include, for example, two completely different antennas or one antenna rotated at some angle ($\theta_{\rm rot}$) with respect to the other. We analyze the following four cases in terms of constraining the foregrounds and the cosmological 21-cm signal, and compare with the current configuration of REACH.
\begin{enumerate}
    \item Dipole + Dipole ($\theta_{\rm rot}=90^\circ$) with $n_{\rm LST}=1$
    \item Dipole + Dipole ($\theta_{\rm rot}=90^\circ$) with $n_{\rm LST}=6$
    \item Dipole + Conical Logspiral with $n_{\rm LST}=1$
    \item Dipole + Conical Logspiral with $n_{\rm LST}=6$
\end{enumerate}

\begin{figure}
    \centering
    \includegraphics[width=\linewidth]{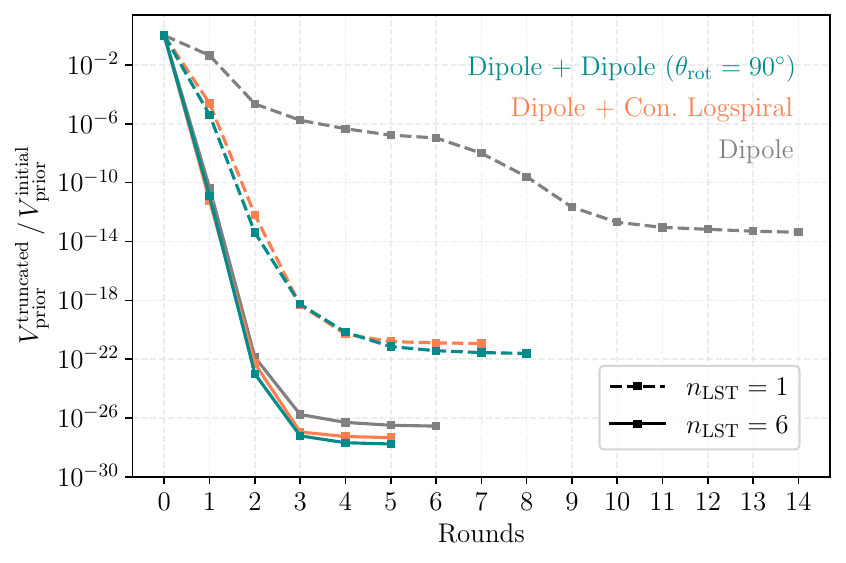}
    \caption{Ratio of truncated prior to initial prior volume for different possible antenna configurations for the upcoming phases of REACH. These are compared with the current configuration of a single dipole antenna with $n_{\rm LST}=1$ and $n_{\rm LST}=6$.}
    \label{fig:truncVol_future}
\end{figure}

In each case, we apply the {\small TMNRE} algorithm, where we concatenate the data from different antennas and time slices, which is given as input to the MLP. In a more traditional approach, one has to work with a beam and time-dependent likelihood function. In Figure~\ref{fig:truncVol_future}, we show the evolution of the ratio of truncated prior volume to initial prior volume for each case until the algorithm converges. These possible configurations are compared with the current configuration of REACH shown in the black line. We note that for a fair comparison, the total integration time is 12 hours for each case, i.e., \ for multiple antennae (and time) analysis, it is uniformly distributed for each antenna (and time). From the figure, we notice that for a single time slice ($n_{\rm LST}=1$), adding another dipole antenna rotated at 90$^\circ$ (cyan) or a conical log spiral antenna (coral) to the existing setup of REACH reduces the truncated prior to initial prior volume by seven orders of magnitude resulting in improved parameter constraints. This improvement comes from the fact that the beam-weighted foregrounds are different for different antennas; however, the 21-cm signal remains the same \citep{10.1093/mnras/stad1047, 10.1093/mnras/stad156}. 

In the case of multiple time slices ($n_{\rm LST}=6$), we notice an order of magnitude reduction in the truncated prior volume by adding a different antenna in the analysis. This improvement is not as significant compared to a single time slice because a significant co-variance between the foregrounds and the 21-cm signal has already been reduced by simultaneously fitting multiple time slices. Another interesting thing to note is that for multiple antennas (and time slices), the algorithm converges faster than a single antenna (and a single time slice). This is achieved because the network sees distinct information corresponding to the same set of parameters. 

%% file: sections/summary.tex
The data analysis of any sky-averaged 21-cm signal experiment generally involves high dimensional parameter spaces to properly take into account the foregrounds, antenna beam uncertainties, ionospheric effects, receiver noise, and the cosmological signal itself. For such highly complex models, the traditional likelihood-based methods of fitting the observed data could become computationally quite expensive. Moreover, given that the traditional Bayesian fitting, i.e.\ sampling the full joint posterior, typically requires millions of samples, it is not feasible to include physically motivated models of the 21-cm signal in the data analysis.

In this article, we have shown how Simulation-Based Inference ({\small SBI}) provides a step towards evading these issues. We considered simulated mock observations composed of foregrounds, the 21-cm signal, and a radiometric noise component. We then performed {\small SBI} through a technique known as Truncated Marginal Neural Ratio Estimation ({\small TMNRE}) in which neural networks approximate the likelihood-to-evidence ratio. The two key aspects of this approach (i) \textit{marginalization} i.e.\ directly estimating the marginal posteriors and (ii) \textit{truncation} of wide priors in multiple rounds based on the posterior densities at the end of each round, significantly reduce the simulation budget and leads to a targeted inference. These two features make our approach computationally more efficient and scalable to high-dimensional parameter spaces. 

We first demonstrated the foregrounds and 21-cm signal reconstruction for the current setup of the REACH experiment, which is observing the sky with a hexagonal dipole antenna. We first considered a Gaussian 21-cm signal model, which is generally used as an approximation for the cosmological 21-cm signal. We found that the network first learns the foreground parameters, truncates their prior distribution, and later identifies the 21-cm signal. However, the constraints on some of the 21-cm signal parameters are found to be weak when we analyzed the data from a single time slice $(n_{\rm LST}=1)$ of 12 hours. This was the consequence of a large co-variance between the beam-weighted foregrounds and the 21-cm signal. This overlap is significantly reduced by analyzing the data from multiple time slices and utilizing the fact that the beam-weighted foregrounds but not the 21-cm signal would be different for different time slices. A simultaneous fitting of multiple time slices ($n_{\rm LST}=6$) reduces the truncated prior volume by $\sim12$ orders of magnitude in comparison to $n_{\rm LST}=1$.

We further presented a validation of our inference results through a statistical coverage test by quantifying the nominal and expected coverage probabilities of our trained network. We found excellent posterior coverage, indicating that the network is neither conservative nor over-confident and that we have a well-calibrated posterior distribution of model parameters. We then applied our algorithm to a physically motivated 21-cm signal model, which is computationally rather expensive to explore within a likelihood-based framework. However, the efficiency of the {\small SBI} approaches allows one to use such complicated models. This is crucial because the resulting parameter constraints can be directly linked to the astrophysics of the early Universe.

Finally, we discussed potential paths for the REACH experiment in Phase II, where we applied our algorithm to several different configurations involving a second observing antenna in addition to the current hexagonal dipole. This also leads to an improvement in the parameter constraints due to the beam-weighted foregrounds being different for different instruments but the 21-cm signal being the same. We found that simultaneously analyzing the data from multiple antennas and time slices would maximize the information content and thus lead to a better search for the cosmological signal. 

The signal extraction, as presented in this work, could be further improved with the inclusion of instruments sensitive to the polarization of the sky radiation to reduce further the co-variance between the foregrounds and cosmological signal \citep{2018ApJ...853..187T}. Our approach is flexible enough to marginalize over nuisance parameters, which describe the ionospheric distortions, {\small RFI}, errors in the foreground map \citep{10.1093/mnras/stad3392} and beam uncertainties \citep{cumner2023effects}. These effects will be considered in future work.